\documentclass[epjCONF, onecolumn]{svjour}
\usepackage{graphics}
\usepackage[varg]{txfonts} 
\usepackage[latin1]{inputenc}
\session-title{Assembling the puzzle of the Milky Way}
\begin{document}
\title{Could M31 be the result of a major merger?}
\author{Sylvain Fouquet\inst{1}\fnmsep\thanks{\email{sylvain.fouquet@obspm.fr}} \and François Hammer\inst{1} \and Yang-Bin Yang\inst{1,2}, Jianling Wang\inst{1,2}, Mathieu Puech\inst{1} \and Hector Flores\inst{1}}
\institute{Laboratoire GEPI, Observatoire de Paris, CNRS-UMR8111,
Univ. Paris-Diderot, 5 place Jules Janssen, 92190 Meudon France
\and
NAOC, Chinese Academy of Sciences,
A20 Datun Road, 100012 Beijing, China
}
\abstract{
We investigated a scenario in which M31 could be the remnant of a gas-rich major merger. Galaxy merger simulations, highly constrained by observations, were run using GADGET 2 in order to reproduce M31. We succeeded in reproducing the global shape of M31, the thin disk and the bulge, and in addition some of the main M31 large-scale features, such as the thick disk, the 10kpc ring and the Giant Stream. This lead to a new explanation of the Giant Stream which could be caused by tidal tail stars that have been captured by the galaxy potential.
} 
\maketitle
\section{Introduction}
\label{intro}

Several observations suggest that the Andromeda galaxy could be the remnant of a major merger of two massive metal-rich galaxies: M31 has a robustly classical and not pseudo bulge \cite{Kormendy2010}, the M31 halo has a high metallicity \cite{VdB2005}, and lastly, in the Andromeda outskirts, several streams have been detected showing similar properties in age and metallicity \cite{Ferguson2005}, which suggests a simultaneous formation process. Conversely, the Giant Stream is not easily reproduced by a recent minor merger \cite{Font2008} because it contains no star with age below 5 Gyr, i.e., no recent star formation.
\section{Dating the possible major merger stage}
\label{sec:1}

In order to date the stage of the major merger, i.e., the first and the second passage (almost the fusion time), we constrain merger models by the Giant Stream, halo field and thick disk  star formation. Indeed, the star formation history of a merging system is enhanced from the first passage to the fusion itself \cite{cox08}. Most of the gas and stars are deposited in the remnant outskirts by tidal tails formed at the first passage and later at the fusion of the cores. A few hundred billion years after its formation, the tidal tail density decrease, provoking a natural quenching of the star formation \cite{wetzstein07}. Thus, the age of the material, brought by tidal tails, provides possible dates for both the first passage and the fusion time. In the 21 and 35 kpc fields \cite{brown07} \cite{brown08} assumed to represent the inner halo, there is no significant star formation more recent than 8.5 Gyr. Hence, in the merger model, the first tidal tails would populate these areas and produce the halo enrichment during the first passage from 8.5 to 9 Gyr ago. The thick disk has a star formation history comparable to that of the Giant Stream. Both could be generated by material returning to the galaxy, mostly from tidal tails generated at the fusion. Because their youngest significant population of stars has ages of 5.5 Gyr, the delay between the first passage and the fusion time ranges between 3 and 3.5 Gyr. This could be matched by assuming relatively large impact parameters (24-28 kpc are used in the simulations).
\section{Simulation model}
\label{sec:2}

We used the GADGET2 hydrodynamical code \cite{springel05} supplemented by star formation, feedback and cooling prescriptions \cite{cox08} \cite{wang11}. Low resolution ($\sim 10^5$ particles) was used to simulate the large scale aspects, high for reproducing the faint Giant Stream ($ < 5 \times 10^5$). Observations strongly constrain the initial conditions of the simulations. According to \cite{hopkins09}, a mass ratio from $\sim$ 3:1 to $\sim$ 2:1 is required to produce a remnant bulge with B/T $\sim$ 0.3, like for M31. The total baryonic mass in the simulation must be close to the M31 one, $1.1 \times 10^{11}$ M$_{\odot}$. A prograde-retrograde orientation for the spin axis is favourable in order to rebuild the disk \cite{hopkins08}. In order to reproduce the 10 kpc ring, we choose a polar orbit. Reforming an Sb-like thin disk calls for high gas fraction just before the fusion, i.e, more than 50\%. Therefore, low star formation and gas-rich progenitors, with more than 65\% of gas, are used in order to keep enough gas before the fusion. These assumptions have theoretical grounds. It has been estimated that many high-z galaxies are gas rich and could be the M31 progenitors. The low star formation prior to the fusion could be caused by a possible combination of several effects. First one is a high feedback efficiency in primordial medium. The gas in the progenitors is less concentrated than in present-day spirals as for the low-surface brightness galaxies. The cooling is less efficient in a relatively pristine medium. Lastly, the expected increase of the gas metal abundance (expected slow before the fusion, very efficient during the fusion) may help to increase the molecular gas fraction, the optical depth of the gas, and the radiation pressure effects, all contributing to a change in the star formation history during the interaction (T. J. Cox 2010, private communication). The last constraint, the Giant Stream, is reproduced by tidal tail stars that have been captured by the galaxy potential. The formation of this structure is aligned along the trajectory of the satellite which falls into the mass center at the fusion, 4.5 Gyr after the beginning of the simulation.
\section{Conclusion}

In this study, we develop and test the idea that the M31 could be the result of a gas-rich major merger. Thanks to the M31 star formation, we constrain the dates of the different merger stages, and morphological observations show several large-scale features, which constrain the initial conditions of the galaxy properties and their orbit. Finally, we are able to reproduce the M31 galaxy and some of its faint structures, like the Giant Stream, the thick disk and the 10kpc ring.
\end{document}